\documentclass[pre,aps,preprint, superscriptaddress]{revtex4-1}
\pdfoutput=1
\usepackage{amsmath,natbib,amsfonts}
\usepackage{amssymb,color, float}
\usepackage{graphicx}
\usepackage[centerlast]{caption}
\usepackage{subfig}
\usepackage{tikz}
\usepackage{pgfplots}
\usepackage[colorlinks=true, citecolor=blue, urlcolor = blue, linkcolor= red,bookmarks=true]{hyperref}
\usepackage{epstopdf}
\graphicspath{{figures/}}
\newcommand{\beqa}{\begin{eqnarray}}
\newcommand{\eeqa}{\end{eqnarray}}

\begin{document}
\title{Reentrant Melting of Lanes of Rough Circular Discs}
\author{Md. Samsuzzaman} 
\author{A. Sayeed} 
\affiliation{Department of Physics, Savitribai Phule Pune University, Pune-411007, India.}
\author{Arnab Saha}
\affiliation{Department of Physics, University Of Calcutta, 92 Acharya Prafulla Chandra Road, Kolkata-700009, India}
\email{sahaarn@gmail.com}
\date{\today}

\begin{abstract}
We consider binary suspension of rough, circular particles in two dimensions under athermal conditions. The mean density of the system is kept constant. The suspension is subject to a time-independent external drive in response to which one half of the particles are pulled along the field direction whereas the other half is pushed in the opposite direction. Simulating the system with different magnitude of external drive in steady state, we obtain oppositely moving macroscopic lanes only for a moderate range of external drive. Below as well as above the range we obtain states with no lane. Hence we find that no-lane state re-enters along the axis of the external drive in the non-equilibrium phase diagram corresponding to the laning transition, with varying roughness of individual particles and external drive. Inter-particle friction (contact dissipation) due to the roughness of the individual particle is the main player behind the re-entrance of no-lane state at high external drives.   
\end{abstract}
\maketitle  

\section{Introduction}
The relation between the microscopic processes and macroscopic behavior of a system is pertinent to many scientific discourse. Lane formation is a non-equilibrium self-organisation process observed in varied binary mixtures of two oppositely moving species. Under certain conditions, the microscopic elements of these two species segregate from a homogeneously mixed state to form oppositely moving macroscopic lanes composed of different species. This dynamical transition occurs in widely varying time and length scales starting from pedestrian dynamics \cite{helbing2005self, helbing1998self}, army ants \cite{couzin2003self}, all the way to collective cell-dynamics\cite{yamao2011multi, kogler2015lane, mccandlish2012spontaneous}, driven binary plasma \cite{surko1989positron, sutterlin2009dynamics}, granular flows \cite{aranson2006patterns, borzsonyi2009patterns, mullin2000coarsening} and driven colloids \cite{leunissen2005ionic, vissers2011lane, vissers2011band}.  \\ 

Colloids, when driven far from thermal equilibrium, exhibit diverse spatio-temporal patterns and novel transport properties due to  intricate self-organization processes \cite{yethiraj2007tunable, dobnikar2013emergent}. One of such prototypes of non-equilibrium phenomena in driven binary colloids is lane formation. It has been studied extensively in theory \cite{dzubiella2002lane, lowen2003nonequilibrium, klymko2016microscopic, wachtler2016lane, rex2007lane, rex2008influence, rex2008kolloidale} and in experiments \cite{leunissen2005ionic, vissers2011lane, vissers2011band}. In this particular problem a homogeneously mixed binary colloid is considered where the constituent colloidal particles of different species move in opposite directions in response to an external drive. It has been shown that depending on various parameters of the system (e.g. average density, strength of external drive, temperature etc.), oppositely moving lanes composed of the particles of different species present in the suspension, can emerge from a homogeneously mixed state both by constant (i.e. time-independent) \cite{leunissen2005ionic, vissers2011lane} as well as time-periodic (e.g. ac drive) external drive \cite{vissers2011band}. The colloidal lanes can appear perpendicular \cite{vissers2011band} as well as parallel \cite{vissers2011lane} to the external drive. In case of the lane formation parallel to the external drive the transition occurs when the magnitude of the external driving force exceeds a finite threshold value\cite{dzubiella2002lane}. As the lane forms, it has been shown that the correlation decays algebraically parallel to the drive and exponentially perpendicular to the drive \cite{poncet2017universal}. The transition is first order \cite{dzubiella2002lane} and has considerable finite size effects \cite{glanz2012nature}. For example, in finite system though the transition is quite sharp, in thermodynamic limit when the system-size diverges, the correlation along the external drive does not, signifying a smooth crossover in this limit \cite{glanz2012nature}. Another major finding is the re-entrance of no-lane state with increasing density and keeping other parameters (e.g. external force) fixed.  For a fixed driving force (high enough to form lanes) with increasing particle density, first there occurs a transition from the no-lane state towards the laned state which is followed by a second transition which brings the no-lane state back as the particles get jammed at high densities \cite{chakrabarti2004reentrance}. \\

Recently it has been realised that microscopic frictional contacts among the colloidal particles induced by surface roughness of the particles can produce novel macroscopic effects. An important example of such tribological effect is discontinuous shear thickening (DST) \cite{williamson2002some, cheng2011imaging, barnes1989shear, mewis2012colloidal, brown2009dynamic, hsu2018roughness, hsiao2019experimental} where effective viscosity of the suspension increases abruptly by several orders of magnitude (i.e. much stronger dependence than Bagnold scaling \cite{bagnold1954experiments}) with increasing shear rate  close to its critical value. DST is exhibited in a variety of suspensions with Brownian \cite{metzner1958flow, hoffman1972discontinuous, hoffman1974discontinuous, bender1996reversible, frith1996shear, fagan1997rheology, shrivastav2016heterogeneous, boersma1990shear, d1993scattering, o2000stress, maranzano2001effects, maranzano2001effectsof} as well as non-Brownian \cite{boersma1990shear, lootens2004gelation, lootens2005dilatant, larsen2010elasticity, bertrand2002shear, brown2009dynamic, brown2012role, fall2010shear, fall2012shear, lemaitre2009dry} characters suggesting DST as a universal behaviour of dense suspensions where thermal motion of the particles does not seem to contribute significantly (e.g. \cite{lemaitre2009dry}). Indeed, DST occurs in experiments with rigid non-Brownian neutrally buoyant particles suspended in a Newtonian fluid in the Stokes regime \cite{brown2009dynamic, brown2012role}. Here we also note that depending on the density and surface roughness of the particles, the inter-particle friction can also be pivotal in the collective dynamics of {\it{active}} systems such as pedestrian dynamics or in the dynamics of army ants. \\

Being motivated by above findings, we will explore here how the lane formation qualifies in the presence of inter-particle friction under athermal conditions. In order to introduce dissipative inter-particle force we will adapt a simple route introduced to model mechanics of foam \cite{durian1995foam}  which later become useful to explore other related problems: shear thinning in adhesive dispersions \cite{irani2019discontinuous}, dissipation and rheology of soft-core discs under shear \cite{vaagberg2014dissipation}, plasticity in sheared glass\cite{varnik2014correlations}, avalanche-size distribution in sheared amorphous solid under athermal condition \cite{lemaitre2009rate, lemaitre2009dry}, influence of attractive interaction in granular suspensions \cite{irani2014impact}, jamming in confined, soft particles under gravity \cite{chaudhuri2012dynamical}, effective temperature in driven systems \cite{o2004effective} etc. Essentially it assumes that the inter-particle dissipative force that drags a particle, is proportional to the velocity of the particle relative to the velocities of its nearest neighbours which are in contact to the particle. In the presence of such inter-particle friction, here we will show that for a fixed density, in two dimensions, under athermal conditions, the lanes, which are formed as the external drive exceeds a threshold value (as also shown by the earlier studies on colloidal lane formation \cite{dzubiella2002lane}), will become unstable if the external drive is increased further. In other words, with finite inter-particle friction, at fixed average density but with increasing strength of external drive, there will be a re-entrance of no-lane state. Here we emphasize that the re-entrance of the no-lane state obtained here differs fundamentally from the colloidal re-entrance obtained earlier in \cite{chakrabarti2004reentrance} as it was obtained at fixed external drive and with increasing average density of the system where jamming played a crucial role. \\

Next we will systematically detail our findings, starting from model description and then by describing and analysing the results obtained by simulating the model. For better understanding, at the end we come up with an intuitive phenomenology behind the re-entrance by which we propose a simpler toy model producing qualitatively similar results as obtained by simulating actual model equation. 




\section{Model}

We consider a binary suspension of `a' and `b' type particles in two dimensions where each of the components has $N_a$ and $N_b$ particles within area $A$. The average density of the system $\rho=(N_a+N_b)/A$. The average densities of each of the components are $\rho_a=N_a/A$ and $\rho_b=N_b/A$. Here we consider the $1:1$ mixture where $N_a=N_b=N$.

The particles are interacting with each other via a conservative force derived from an effective pair potential $U$ that depends on the distance between the particles of the concerned pair. For simplicity we consider the symmetric case where $U_{aa}=U_{bb}=U_{ab}=U$.  The inter-particle interaction that we consider here is screened Coulomb interaction \cite{dzubiella2002lane},   

\begin{equation}
U(r_{ij})=V_0\frac{\exp(-\kappa (r_{ij}-\sigma))}{(r_{ij}/\sigma)}.
\label{Yukawa}
\end{equation}
Here $r_{ij}$ is the distance between a pair of the particles denoted by the index $i$ and $j$, $V_0$ is the energy scale and $\sigma$ is the particle diameter that sets a length scale. The inverse screening length $\kappa=4\sigma$ governs the range of the interaction. This interaction is used to model charge-stabilised suspensions where $\kappa$ is the range of the interaction which can be tuned, for example, chemically\cite{crocker1994microscopic}.

Apart from the conservative inter-particle forces, the particles are also going through dissipative forces. One of them is originated from the frictional drag force ${\bf{F}}_i^p$ due to the surrounding fluid. It is proportional to velocity ${\bf{v}}_i$ of the $i-$th particle, i.e. ${\bf{F}}_i^p=-\gamma {\bf{v}}_i$ where $\gamma=3\pi\eta\sigma$ \cite{dhont1996introduction} is the friction coefficient between the particle and the fluid that depends on the viscosity $\eta$ of the surrounding fluid. Here we assume that the effect of  hydrodynamic interaction among the suspended particles are negligibly small which can be due to screening \cite{riese2000effective}.  


Another source of dissipative force that an individual particle can face, is from inter-particle friction. It is a short-ranged dissipative force that depends on the average relative velocity between a particle and its nearest neighbours which are in contact \cite{o2004effective, lemaitre2009rate, lemaitre2009dry, vaagberg2014dissipation} i.e., ${\bf {F}}_i^q=-\frac{\mu}{N_i}\sum_jH(u_{ij})({\bf {v}}_i-{\bf{v}}_j)$ where $H(u_{ij})=1$ if $u_{ij}=\sigma-r_{ij} \geq 0$ and $H(u_{ij})=0$ otherwise. Here $N_i$ is the number of nearest neighbours of $i$th particle for which $H=1$. If $N_i=0$, as there is no neighbouring particle, ${\bf {F}}_i^q=0$.   

Finally, the particles are also experiencing constant external field along a particular direction (here without loosing any generality we choose the direction to be $\hat{\bf{x}}$). Type `a' particles are driven along $+\hat{\bf{x}}$ and `b' type particles are driven in opposite direction due to the external force ${\bf{F}}_{\text{ext}}=F_{\text{ext}}\hat{\bf{x}}_i$.  Clearly, it is only the response to the external force that distinguishes between `a' and `b' type particles. Therefore the equation of motion of the particles with unit mass is given by $\frac{d^2{\bf {r}}_i}{dt^2}={\bf{F}}^p_i + {\bf{F}}^q_i -\sum_j\nabla_iU \pm {\bf{F}}_{\text{ext}}\label{eom}$
where the positive sign is used when $i\in N_a$ and negative sign is used when $i\in N_b$. Thermal fluctuation is considered to be negligibly small in comparison to other forces as mentioned before. 

We consider $\sigma$ to be the unit for length and $1/\gamma$ to be the unit for time. With these units the left hand side of the equation of motion mentioned before, becomes $\gamma^2\sigma\frac{d^2{\tilde{\bf {r}}_i}}{d\tilde t^2}$ where $\tilde{\bf{r}}_i$ is dimensionless position vector of the particle at dimensionless time $\tilde t$. With these units the frictional drag from fluid becomes $\gamma^2\sigma{\tilde{\bf{F}}}^p_i$ where ${\tilde{\bf{F}}}^p_i$ is dimensionless velocity of $i$th particle. Similarly inter-particle friction becomes $\mu\gamma\sigma{\tilde{\bf{F}}}^q_i=\alpha\gamma^2\sigma{\tilde{\bf{F}}}^q_i$ where $\mu/\gamma = \alpha$ and ${\tilde{\bf{F}}}^q_i$ is the dimensionless relative velocity between a particle and its nearest neighbours which are in contact. By dividing the both side of the equation of motion (mentioned earlier) with $\gamma^2\sigma$ and introducing dimensionless gradient operator as $\frac{1}{\sigma}\tilde\nabla_i$ we obtain dimensionless inter-particle energy scale as $U_0=V_0/\gamma^2\sigma^2$. Similarly we obtain  non-dimensionalised magnitude of the externally applied driving force as $\tilde F_{\text{ext}}=F_{\text{ext}}/\gamma^2\sigma$. Therefore the equation of motion with dimensionless quantities reads 

\begin{equation}
\frac{d^2{\tilde{\bf {r}}}_i}{d\tilde t^2}={\tilde{\bf{F}}}^p_i + \alpha{\tilde{\bf{F}}}^q_i -\sum_j\tilde\nabla_i \tilde U \pm {\tilde{\bf{F}}}_{\text{ext}}.
\label{eom1}
\end{equation}  
where $\tilde U=U_0\frac{\exp(-\kappa\sigma (\tilde{r}_{ij}-1))}{\tilde{r}_{ij}}$, $\tilde r_{ij}=r_{ij}/\sigma$. The average non-dimensional density is given by $\tilde\rho =\rho\sigma^2$. For  brevity of notation, from now on, we will omit tilde from the dimensionless variables.  

\section{Simulation Method}
We have carried out simulation by integrating the equation of motion given in Eq.[\ref{eom1}]. The data presented here are obtained by simulating $N=2000$ particles confined within a square box of area $A = 44\sigma\times44\sigma$ with periodic boundary condition in both X and Y directions. The average density is suitable for the lane formation provided the external driving force exceeds a threshold value \cite{chakrabarti2004reentrance,dzubiella2002lane}.  To update the position and velocity of the particles we use velocity verlet algorithm \cite{allen2017computer} with time discretization $\delta t=0.0001$, for a total simulation run of $6\times10^7$ iterations. Positions and velocities of the particles are recorded when the system reaches at non-equilibrium steady state after $50\times 10^4$ simulation steps when the average lane order of the system does not alter significantly over time. Initially the positions and velocities of the particles are distributed randomly. Once ${\bf F}_{\text{ext}}$  is switched on, $N_a$ and $N_b$ particles are chosen randomly from $2N$ particles such that the initial configurations represent well-mixed binary mixture of $a$ and $b$ type particles. For all our simulations we have kept the values of $A, N$ (\text{and thereby}  ${\rho}), \gamma, V_0, \sigma$ constant and vary $|{\bf{F}}_{\text{ext}}|= {F}_{\text{ext}}$ and ${\alpha}$. The results presented here are averaged over time in non-equilibrium steady states and also over 3 realisations. 




\section{Results}

We begin by analyzing the non-equilibrium steady states of the system obtained by varying $F_{\text{ext}}$ and $\alpha$. The states are characterised by the spatial organisation as well as the dynamical properties of the particles. Below we will develop the tools to characterise the steady states.  

In a similar set-up with colloidal binary suspensions, it is known that the system phase-separate at finite temperature and  the components move in opposite directions forming oppositely moving parallel lanes \cite{dzubiella2002lane}. The transition from non-laned state to laned state in colloidal binary suspension starts to occur after a finite threshold of external driving force. The transition is reversible and exhibits significant hysteresis. It is classified as a first-order non-equilibrium transition \cite{dzubiella2002lane}. Note that there are fundamental differences between the system that we are concerned here and the system discussed in \cite{dzubiella2002lane} : the present model includes inter-particle contact dissipation and it does not involve thermal fluctuation which is relevant for non-Brownian suspensions ( e.g. \cite{o2004effective}). Our aim here is to explore the consequence of inter-particle friction on lane formation in two dimensions when the thermal fluctuation is negligibly small. To characterise the spatial organisation of the particle in the context of lane formation, we borrow the lane order parameter from \cite{dzubiella2002lane}. We define the lane order parameter by assigning every particle a quantity $\phi_i$ as,


\beqa
\nonumber
\phi_i &=& 1 ~~~ {\text{when}}~~ |y_j-y_i|>\rho^{-1/2}/2\\
&=& 0 ~~~ {\text{elsewhere}}          
\label{local_lane_order}
\eeqa 
where $i\in N_a, j\in N_b$. This can be termed as local (i.e. defined for individual particle at time $t$) lane order parameter. We define the global (i.e. defined for the whole system at a given configuration at time $t$) lane order parameter as,
 
\begin{equation}
\phi=\frac{1}{N}\sum_i\phi_i
\label{global_lane_order}
\end{equation} 
which can then be averaged over time and realisations in steady states. Note that for a perfectly mixed, non-laned state $\phi$ is zero. It increases from zero when the lanes start to appear. For a perfectly laned state $\phi$ is unity.  

We will see later that with increasing $F_{\text{ext}}$, after a certain threshold value (say $F_{c}$), particles phase separate forming oppositely moving lanes. This phenomena is similar to the binary colloidal suspension as in \cite{dzubiella2002lane}. If $F_{\text{ext}}$ is increased further the laned state continues up to a second threshold value of $F_{\text{ext}}$ (say, $F_{c}^\prime$). If $F_{\text{ext}}$ is increased beyond $F_{c}^{\prime}$, irrespective of their type, particles tend to accumulate randomly creating high as well as low density regimes. Hence the density profile of the system tends to be increasingly non-uniform with increasing $F_{\text{ext}}$. Eventually the lane structure is broken at high enough $F_{\text{ext}}$  and we obtain the reentrance of non-laned state, which is the central theme of the present work. Therefore it is important to quantify the non-uniformity of the density distribution of the system. We will quantify the non-uniformity by calculating the standard deviation of the local density profile $g(x,y)$ of the system. Here $g(x,y)$ is the local probability density of finding a particle (irrespective of its type) between $x$ to $x+dx$ and $y$ to $y+dy$.  Clearly $\int g(x,y)dxdy=2N$. Computationally $g$ is obtained by dividing the system into a number of cells and then by counting the number of particles  in each cell. For a given configuration the standard deviation is given in Eq.\ref{sigma} where the angular bracket represents averaging over number of cells. It is noteworthy that the cell should not be very large as it becomes incapable to capture the non-uniformity of the density profile (extreme case :  for the cell which is of the same size as the system, $\Sigma$ vanishes).  It should not be very small as well i.e. it should be larger than the size of a single particle. The cell-size we choose here is $4\sigma^2$. Here we report $\Sigma$ after taking the averages over time and realisations in steady states. 
\begin{equation}
\Sigma=\sqrt{\langle(g(x,y)-\rho)^2\rangle}
\label{sigma}
\end{equation}

We emphasize here that the types of the particles are not considered while calculating $g(x,y)$. Hence, for the states with uniform distribution of particles,  irrespective of whether they contain lanes or not, $\Sigma$ is very low. In other words, for states with uniform particle distribution, by measuring $\Sigma$ one cannot distinguish between laned and non-laned states. $\Sigma$ increases when particles, disregarding their types, get accumulated in random places, making the density profile non-uniform. $\Sigma$ is a measure of the non-uniformity in the position distribution of the particles. 
 


\begin{figure}[H]
\centering
\includegraphics[width=12cm,height=12cm,angle=0]{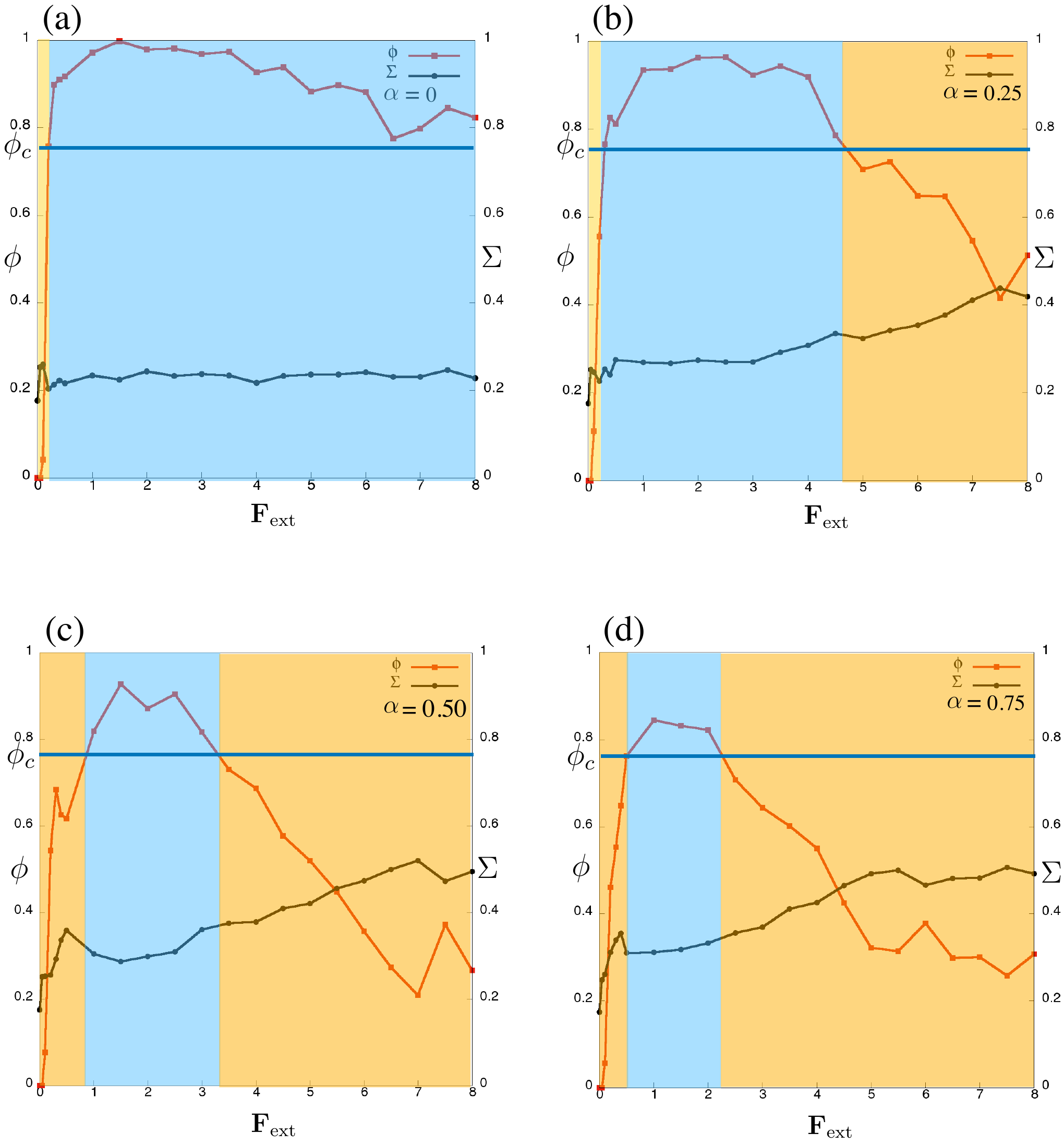}

\caption{Lane order parameter $\phi$ and standard deviation $\Sigma$ of the density profile are plotted with external drive $F_{\text{ext}}$  for $\alpha = 0,0.25,0.5,0.75$ in (a), (b), (c), (d) panels respectively. $\phi$ is in red and $\Sigma$ is in black. The threshold of the lane order $\phi_c$ is shown by $\phi=\phi_c$ line in blue. The width of the blue region depicts the window $\Delta F=F_c^{\prime}-F_c$ within which $\phi > \phi_c$ i.e. quality lanes occur in the system. The yellow region is where $\phi < \phi_c$. $F_{\text{ext}}=F_c$ is the left boundary of the blue region, which is the first threshold value of $F_{\text{ext}}$ beyond which quality lanes form in steady states and therefore $\phi$ becomes larger than $\phi_c$. With finite inter-particle friction (i.e. $\alpha > 0$ ), $F_{\text{ext}}=F_c^{\prime}$  is the right boundary of the blue region, which is the second threshold value of $F_{\text{ext}}$  beyond which, lane quality drops and $\phi$ becomes smaller than $\phi_c$. Clearly, without inter-particle friction (i.e. for $\alpha=0$ ), once the lanes are formed their quality does not drop considerably with increasing $F_{\text{ext}}$ and therefore $\phi$ is larger than $\phi_c$ for all values of $F_{\text{ext}}$ considered here. It is apparent from panels (b) - (d) that for $\alpha > 0$, $\Sigma$ increases with $F_{\text{ext}}$, affecting the lane-quality. Hence $\phi$ drops below $\phi_c$ when $F_{\text{ext}} > F_c^{\prime}$. Hence the no-lane state, which was there for $F_{\text{ext}} < F_c$, re-enters when $F_{\text{ext}} > F_c^{\prime}$.}
\label{laneorder}
\end{figure}
The variation of $\phi$ and $\Sigma$ with external force for different $\alpha$ is plotted in figure \ref{laneorder}. From the plots it is evident that with $\alpha>0$ ,  $\phi$ has non-monotonic dependence on $F_{\text{ext}}$ .  Initially when $F_{\text{ext}}=0$, the $a$ and $b$ type particles are homogeneously mixed with each other and therefore it is a no-laned state with $\phi=0$. It increases with increasing $F_{\text{ext}}$  and reaches at a plateau when $F_{\text{ext}}$  goes beyond the threshold $F_c$ . In this plateau region even if $F_{\text{ext}}$ increases, $\phi$  does not change considerably. This plateau continues till $F_{\text{ext}}$ reaches at a second threshold value $F_c^{\prime}$ . When $F_{\text{ext}}>F_c^{\prime}$ , $\phi$ starts to decrease considerably with increasing $F_{\text{ext}}$. Hence the non-laned state re-enters. To distinguish between laned and no-lane state we consider a threshold $\phi_c=0.75$ such that for laned states $\phi>\phi_c$  and for no-lane states $\phi<\phi_c$. We note from the figure that for $\alpha=0$, $\phi>\phi_c \forall F_{\text{ext}}>F_c$  which is the reminiscent of lane formation in binary colloidal suspensions as explored in \cite{dzubiella2002lane}. For $\alpha>0$, due to the re-entrance of no-lane state, $\phi>\phi_c$  for $F_c<F_{\text{ext}}<F_c^{\prime}$ . Beyond this window of external drives $\Delta F=F_c^{\prime}-F_c$, $\phi<\phi_c$. It is evident from the figure that $\Delta F$ decreases as $\alpha$ increases. 
In contrast to $\phi$, when $\alpha=0$, $\Sigma$ does not change considerably with $F_{\text{ext}}$. This implies that though the particles of different species segregate to form lanes, irrespective of their species, they are uniformly distributed through out the system. On the other hand when $\alpha > 0$, in contrast to $\phi$, $\Sigma$ increases monotonically with $F_{\text{ext}}$. This signifies that in presence of inter-particle friction, with increasing  $F_{\text{ext}}$, the density profile becomes increasingly non-uniform. Though for lower external drives ($F_c < F_{\text{ext}} < F_c^{\prime}$) this non-uniformity is not so detrimental to the lane-structure of the system such that it falls apart but at higher external forces ($F_{\text{ext}} > F_c^{\prime})$ it is. Hence, with $\alpha > 0$, at high external drives, $\phi$ drops below $\phi_c$ and one gets the no-lane state back.

\section {Phase Diagram}
\begin{figure}[H]
\begin{center}
\includegraphics[width=10cm,height=12.5cm,angle=0]{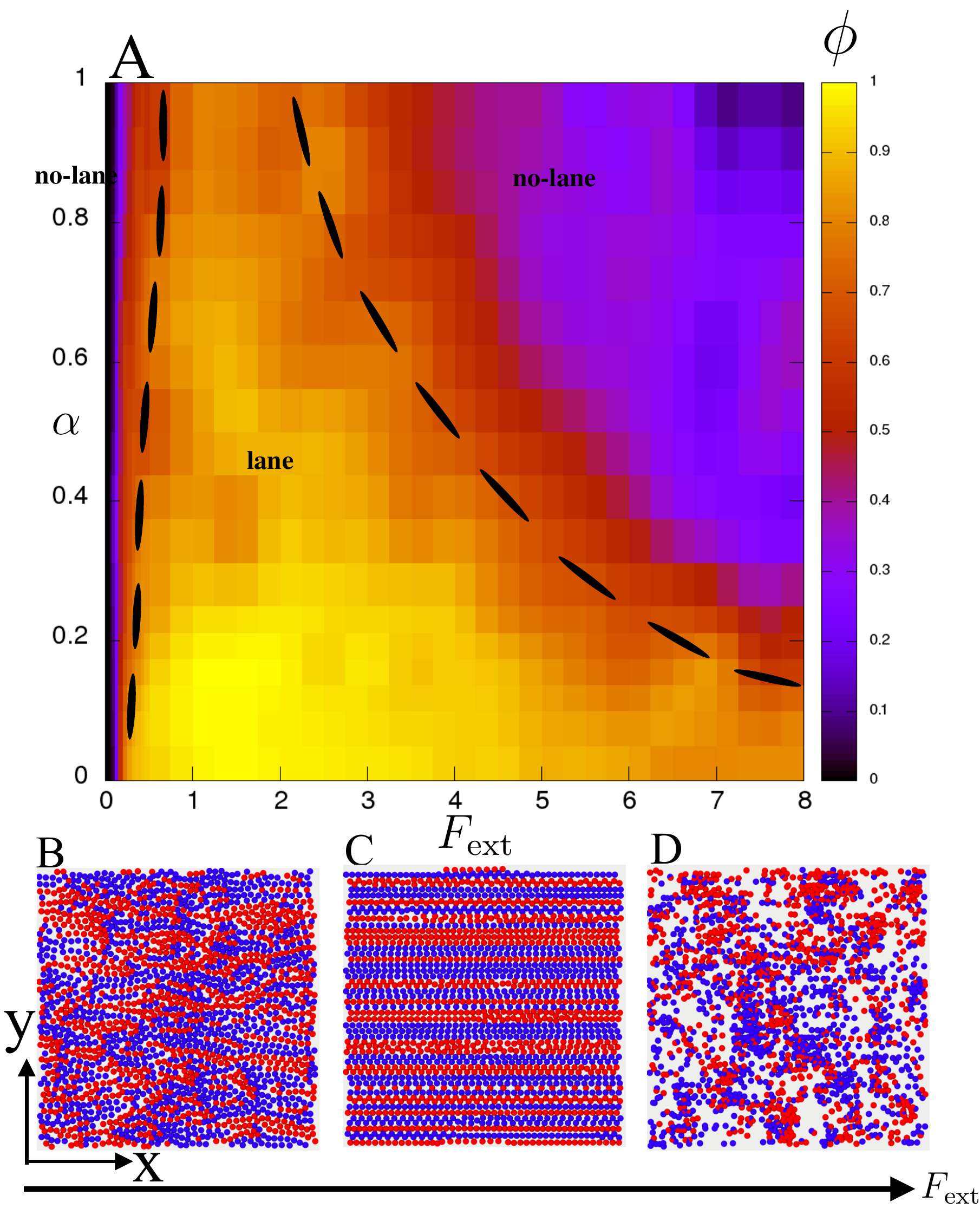}
\caption{A: It is the heat map of the lane order parameter $\phi$ in $F_{\text{ext}}-\alpha$ plane. It indicates that with finite inter-particle friction ($\alpha > 0$), for very low and high $F_{\text{ext}}$, the system does not support any lane structure whereas it is supported by the optimum values of $F_{\text{ext}}$, occur in-between. Broken lines are drawn schematically as a  guide-to-eye to indicate the laned and no-lane phases. The phase diagram is constructed with a total of 22 different values of $F_{ext}$ and 20 different $\alpha$. B--D : Typical configurations with increasing magnitude of external drive $F_{\text{ext}}$, shown by the long black arrow at the bottom. Red and blue colors of the particles indicate that the external drive applied along +$\hat{\bf x}$ and -$\hat{\bf x}$ directions.} 
\label{phasediagram}
\end{center}
\end{figure}

The re-entrant transition between laned and no-lane phases of the system is represented by the phase diagram in $\alpha-F_{\text{ext}}$ plane in figure\ref{phasediagram}. With the heat map of $\phi$ and the broken lines as the guide-to-eye in $\alpha$-$F_{\text{ext}}$ plane we represent the phases. 

For $\alpha=0$ and very low $F_{\text{ext}}$ ($0 < F_{\text{ext}}<0.2$) no significant laning is observed ($\phi <  \phi_c$). As $F_{\text{ext}}$ increases and crosses the threshold $F_c$, system enters into the laned state ($\phi>\phi_c$) and it remains in the laned state for all values of $F_{\text{ext}}$ . There is no signature of re-entrance of the no-lane state with increasing $F_{\text{ext}}$ when $\alpha=0$. This is also true as far as $\alpha$ remains small ($\alpha\lesssim 0.15$). Thus, a minimum external drive is required to segregate the particles into distinct lanes which can be inferred from the fact that the external drive has to be larger than the other forces between any two pair of the particles moving opposite to each other. .

The scenario changes drastically as $\alpha$ increases beyond $0.15$ and the change is apparent for high $F_{\text{ext}}$. In this regime of $\alpha$, as earlier, we observe that lanes appear into the system only when $F_{\text{ext}}>F_c$. $F_c$ increases slightly as we increase $\alpha$. This implies that as the contact dissipation between the particles increases we need higher external drive to move against each other to form lane. The qualitative difference appears as we increase $F_{\text{ext}}$ further. We observe that when $F_{\text{ext}}$ goes beyond a second threshold $F_c^{\prime}$, the lane quality falls. $\phi$ becomes smaller than $\phi_c$. No-lane state re-enters.  Thus, in the $\alpha - F_{\text{ext}}$ plane, along the axis of increasing $F_{\text{ext}}$, first we have a region of no lanes, then a region of lanes and finally another region of no-lane. This is the central theme of our paper, which highlights the fact that it is not possible to have good lanes even if we increase the external drive to very high values, in the presence of inter-particle friction. We also observe that though $F_c$ increases slightly but $F_c^{\prime}$ reduces considerably with increasing $\alpha$. Hence the window $\Delta F$, responsible for laning, shrinks with increasing $\alpha$.  

The typical configurations with increasing $F_{\text{ext}}$ are shown in the lower panels of Fig.{\ref{phasediagram}}, from B to D.  They demonstrate that for very low external drive, in steady state, lanes are not developed within the system (panel B).It is developed only after crossing the finite threshold value of  the drive $F_c$. Typical laned configuration is shown in panel C. When the external drive is increased further and eventually $F_{\text{ext}} > F_c'$, the lanes are broken due to interparticle friction and no-lane state re-enters. Typical steady state configuration in this regime is given in panel D.

\section{Dynamical Property}
The transport along the direction of the external field is affected due to the inter-particle friction. Lanes help transport. With finite inter-particle friction the lane order is reduced at high external drive. Hence it is intuitive that the transport along the external drive will be affected as we increase the field. 
We quantify the transport along the direction of the external field by computing the drift velocity along the field in steady state. The drift velocity is defined as \cite{dzubiella2002lane},
\begin{figure}[H]
\begin{center}
\includegraphics[width=10cm,height=10cm,angle=0]{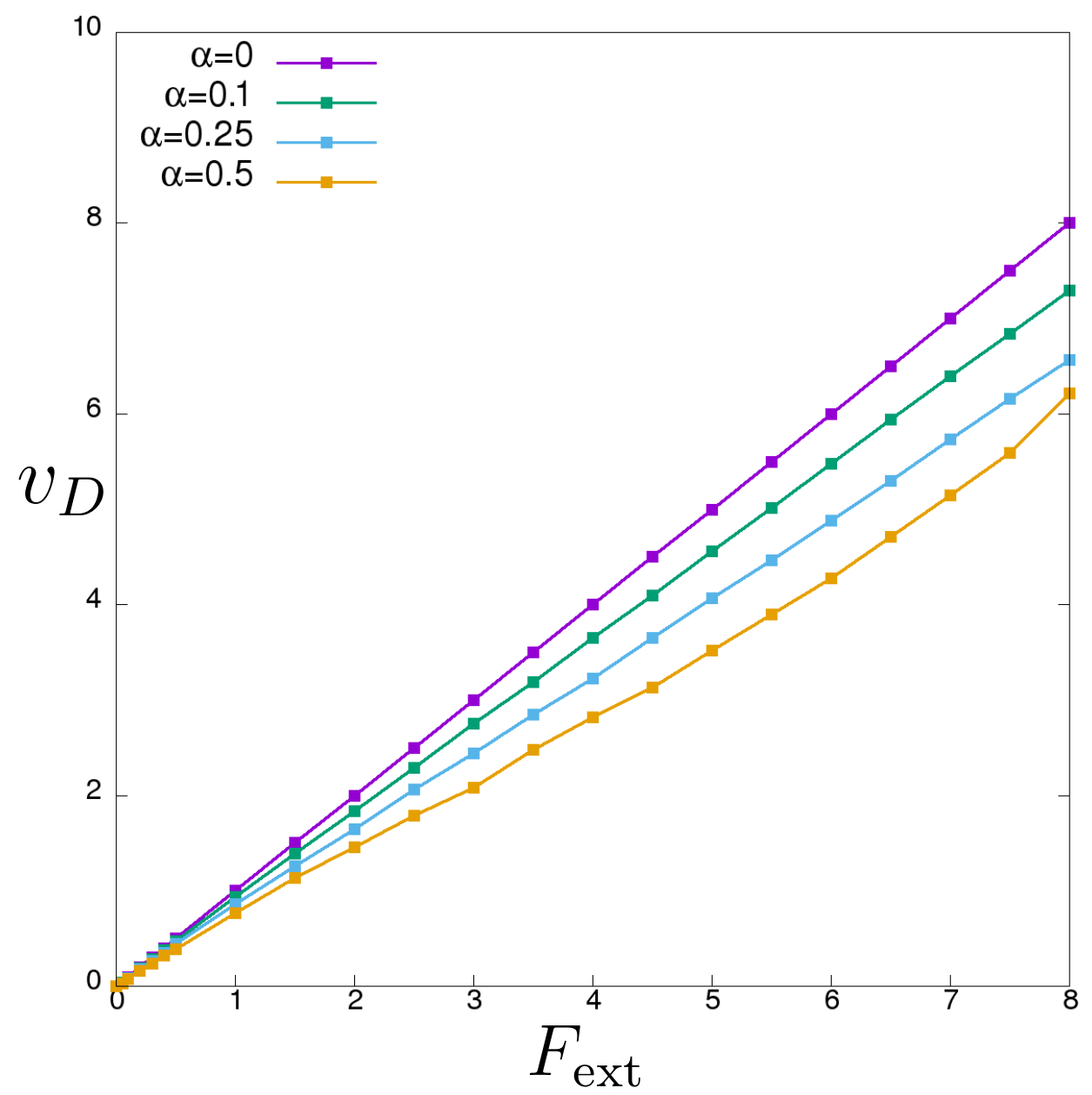}
\caption{Average drift velocity is plotted with different external drive $F_{\text{ext}}$. For every $F_{\text{ext}}$ the drift along the drive is maximum when $\alpha=0$. It reduces gradually with increasing $\alpha$.}
\label{drift}
\end{center}
\end{figure}
\begin{equation}
v_D^{2}=\lim_{t\rightarrow \infty} \frac{\langle(x_i(t) - x_i(0))^2\rangle}{t^2}
\label{drift_vel}
\end{equation} 
where $x_i(t)$ is the $x$-position of $i$-th particle of the system. We measure $v_D$ after the system reaches at a steady state where $\phi$ fluctuates around a constant mean.  The angular bracket implies averaging over different particles and then taking the average over time and realisations in steady states. We observe that as expected, drift velocity along the direction of the external field is reduced with increasing inter-particle friction.

\section{Discussion}

So far we have shown that with increasing external drive, not only the transition from a no-lane state to a laned state occurs in a driven binary suspension of oppositely moving rough particles under athermal conditions but a transition in reverse direction (laned to no-lane state) also occurs when the external drive is large enough. This re-entrant transition occurs when the inter-particle friction due to the roughness of the particles, affects their dynamics significantly.  We have also shown that with increasing external drive, the density profile of the rough ($\alpha > 0$) particles become gradually  nonuniform. Hence the well-formed lanes start to break at high external drives. Here we will discuss how the inter-particle friction due to the roughness of the particles can make the density profile non-uniform at high external drive. 

Consider a state of the system with $F_{\text{ext}}\rightarrow F_c^{\prime -}$ and $\alpha > 0$. In this parameter space, oppositely moving lanes are well-developed and adjacent to each other in the system. Particles which are within a lane and far from the boundaries of the lane, are facing less resistance from its neighbours because all of them are moving along the same direction. For these particles, ${|\bf {F}}^q_i|\simeq 0$. Though for the particles which are at the lane-boundaries they are facing resistance from the oppositely moving particles from the lane next to them. Therefore for them ${|\bf {F}|}^q_i > 0$. Hence the effective friction faced by the interfacial particles between two lanes are more in comparison to the particles which are well inside a lane. This makes the interfacial particles slower in comparison to the other particles of a lane. Eventually it leads the effective friction to be space dependent, making the density distribution non-uniform, particularly at high external forces. 

To test this, below we consider a toy equation of motion for the particles where, instead considering the inter-particle friction explicitly, we consider the effective friction $G > 0$  to be space-dependent.  Considering a configuration of the system with well-developed lanes one may assign properties like $G(x,y)=G(x,y+l)$ and $G(x,l) < G(x,l/2)$ (where $l$ is the typical lane-width) to the effective friction. The inequality ensures that the effective friction inside a lane is less than the friction faced by the particles at the interface between two oppositely moving lanes. 

It is noteworthy that the properties assigned to $G(x,y)$ is relevant only when lanes are well-developed in the system i.e. when $F_c < F_{\text{ext}} < F_c^{\prime}$. In general, for all $F_{\text{ext}}$, $G$ can have complex dependence on position and relative velocity (relative to the neighbours) of a particle, deriving which is beyond the scope of the current paper. Instead we assume a functional form of $G$ that maintains the aforementioned properties for all values external drives. Though the strength of the periodic modulation of $G$ is small enough such that it cannot disturb the lane formation till $F_{\text{ext}}\leq F_c^{\prime}$. When $F_{\text{ext}} > F_c^{\prime}$, we will see that the periodic modulation in $G$ affects the lane structure and eventually the lane order decreases. This essentially indicates that in the presence of contact dissipation, the effective friction faced by a particle within a laned state can be spatially periodic and it can destabilise the lane structure beyond a certain threshold value of external drive $F_c^{\prime}$.

We assume following simple functional form of $G(x,y)$ as,

\begin{equation*}
G(x,y) =\gamma\left(1+a \sin{^2}\left(\pi y/l\right)\right)
\end{equation*}
where $a > 0$ is the amplitude of the periodic modulation of $G$ varying in space between its minimum value at the middle of a lane at $y=nl$, given by $ G_{\text{min}}=\gamma $  and its maximum value at the interface between two oppositely moving lanes i.e. at  $y=(n+1/2)l$, which is given by  $G_{\text{max}}=\gamma(1+a)$ (where$n=0,1,2,3...$). The equation of motion of ith particle will be

\begin{equation}
\frac{d^2{{\bf {r}}}_i}{d t^2}={{\bf{F}}}^p_i  -\sum_j\nabla_i U \pm {{\bf{F}}}_{\text{ext}}.
\label{eom22}
\end{equation}
where the effective friction (both from the fluid and from the neighbouring particles in contact) faced by the particle is given by $ {\tilde{\bf{F}}}^p_i = - G(x,y){\bf {v}}_i$.We simulate the aforementioned model in the same parameter space as before and evaluate $\phi$ with different $|{\bf F}_{\text{ext}}|$ to obtain the following figure \ref{Toy},

\begin{figure}[H]
\begin{center}
\includegraphics[width=10cm,height=10cm,angle=0]{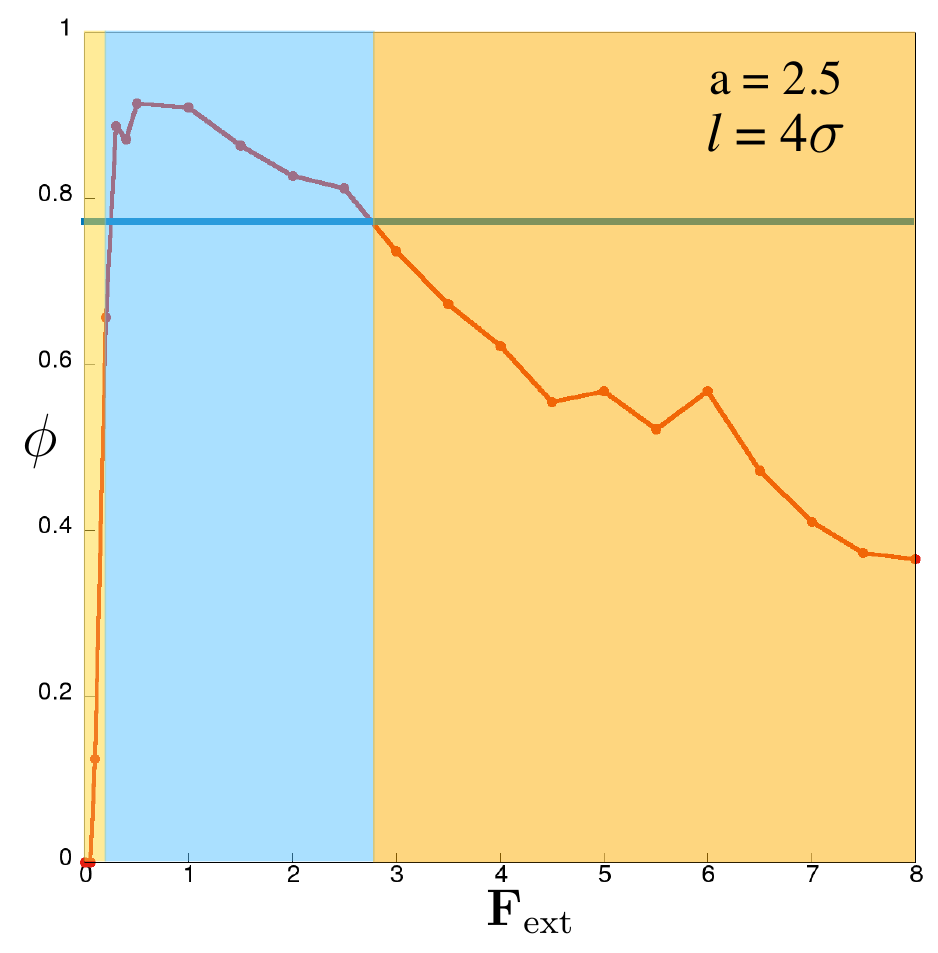}
\caption{Here we have plotted steady state $\phi$ with different external drive $F_{\text{ext}}$ obtained by simulating Eq.{\ref {eom22}}. The blue horizontal straight line represents $\phi=\phi_c$. The blue rectangle represents the optimal values of $F_{\text{ext}}$ for which $\phi > \phi_c$. The left boundary ($F=F_c$) of the rectangle represents the lower threshold of the external drive below which $\phi < \phi_c$. The right boundary ($F=F_c^{\prime}$) of the rectangle represents the upper threshold of the external drive beyond which states with $\phi < \phi_c$ re-enter.}
\label{Toy}
\end{center}
\end{figure}

From the figure it is apparent that the model with space dependent effective friction but without explicit inter-particle friction provides qualitatively same results as obtained in Fig. \ref{laneorder}. Initially, for very small $F_{\text{ext}}$, no lane was observed and $\phi < \phi_c$. Once $F_{\text{ext}}>F_c$ ($F_{\text{ext}}=F_c$ is the left boundary of the blue region in Fig. \ref{Toy}), lanes develop in steady states and hence $\phi > \phi_c$. As we increase $F_{\text{ext}}$ further, once $F_{\text{ext}}>F_c^{\prime}$ ($F_{\text{ext}}=F_c^{\prime}$ is the right boundary of the blue region in Fig.\ref{Toy}), lanes start to break. Eventually $\phi$ becomes less than the threshold $\phi_c$ and no-lane state re-enters. It is important to mention here that though there is qualitative similarity between the results from the aforementioned space-dependent effective friction model (Eq.\ref{eom22}) and the model with explicit  inter-particle friction (Eq. \ref{eom1}) in context to the re-entrant transition between laned and no-lane states, the rigorous relation between these two approaches is not yet established and it is beyond the scope of the current paper.

\section{Conclusion}

We consider a binary suspension of two types of particles in two dimensions, which move in opposite directions when subjected to a constant external drive. The particles are rough and the density of the system is high enough such that the inter-particle friction plays important role to the dynamics of the system. The thermal fluctuations are assumed to be insignificant in comparison to the other forces present in the system. We have shown that when the external drive goes beyond a threshold, the particles of different species segregate from the homogeneously mixed suspension to develop oppositely moving macroscopic lanes. The lane structure continues with increasing external drive up to a second threshold after which inter-particle friction starts to dominate. Consequently the system cannot support the lanes anymore and we obtain a no-lane state. Note that for very low external drive, the system, being homogeneously mixed, does not support lanes. Therefore, with finite inter-particle friction, we obtain the re-entrance of no-lane state along the axis of increasing external drive.   

There are few important related topics which deserve detailed, separate study. Though we leave them out from the current paper but we will mention them here briefly. 

Transformation from no-lane state to laned state in binary colloidal systems (without inter-particle contact dissipation) has significant finite size effect \cite{glanz2012nature}. It important to study the finite size effect here as well particularly for the re-entrance of no-lane state at high external drive. 

From the phase diagram {\ref{phasediagram}} it is apparent that as $\alpha$ increases $F_c$ and $F_c^{\prime}$ come close to each other. Question is, if we increase $\alpha$ further (which is not included in the current phase diagram), will there be a certain $\alpha$ for which $F_c$ and $F_c^{\prime}$ merge to a single point in the phase diagram, beyond which no lane can be formed? If yes, what will be the characteristics of that point in context of the phase transition. 

Last but not least, it will be interesting to study the laning transition in the suspension of rough micro-particles particles in the presence of thermal fluctuations.

We believe our findings here are amenable to experiments with dense non-Brownian systems having significant frictional contacts among the particles (e.g. \cite{hsu2018roughness}). 

\section{Acknowledgement}
Arnab Saha (AS) thanks the start-up grant from UGC, UGCFRP and the Core Research Grant ( CRG/2019/001492) from DST, Govt. of India. AS thanks Gaurav P. Shrivastav and Pinaki Chaudhuri for careful reading and critical comments. 	   

\newpage
\bibliographystyle{plain}
\bibliography{Manuscript_30_08_20_ASNotes}
\end{document}